# Dynamo and Electrical Jet in Hall Plasmas, Application to Astrophysics

L.I. Rudakov

**Visiting Scientist at University of Maryland
and Massachusetts Institute of Technology**

**Abstract.** The magnetic field in Hall plasmas is frozen in the electron component and is advected not only with the plasma motion but also with the electrical current flow. Its coupling with the plasma may be not as strong as characteristic of the MHD approximation. The rotation and slipping of the magnetic field result in a very different and less efficient magnetic field generation by dynamo - rotating plasma disk in magnetic field. We found a some exact analytical solutions of nonlinear equations describing the dynamo. In particular, the dynamo may not dissipate the energy in the steady state limit. The 3-component magnetic field and magnetic energy are generated and accumulated during the transition time only, in the same way as in a nonlinear inductor. An electrical jet is the other unusual phenomenon for MHD. It was investigated theoretically and experimentally for laboratory plasmas during the last 15 years and now is used for fast current switching. We found periodical or shock-like nonlinear wave traveling along a Hall plasma column and not associated with plasma motion. A nonlinear equation describing a possible steady state magnetic field distribution and current flow in Hall plasma conductor is derived that differs from the Grad-Shafranov equation for the low pressure plasma. In conclusion we discuss application of our results to astrophysical plasmas. This physics could be important for understanding the evolution of dusty plasma disks and jets around new stars.

### 1. Introduction

Hall physics appears in quasi-neutral plasmas where one of the plasma components is non-magnetized (for example, holes in semiconductors or laboratory plasma ions on a time scale shorter than the ion cyclotron period) and moving in different way than the magnetized particles. Strong Hall electric field arises then to keep plasma quasi-neutral.
Most of the matter in the interstellar medium consists of gaseous hydrogen. However, about 1 % of mass of matter is in a form of micron and sub-micron size dust particles. The grains are charged and frequently non-magnetized since the drag force acting on the dust particles from the gas flow may be dominant or duration of process is much less then dust cyclotron rotation time.
Astrophysicist take into account charged dust when evaluating magnetic field diffusion in the low temperature weakly ionized molecular clouds during star formation [1,2]. However, there are other less known but possibly more important effects associated with the Hall electric field. The Hall electric field in 2D geometry is not a potential one. As a result, magnetic field may penetrate even in motionless plasma as a magnetic shock wave. This phenomenon was investigated theoretically and experimentally for laboratory plasmas during last 15 years and electrical jet now is used for fast current switching [3]. Hall physics is an important feature in the problem of magnetic field reconnection (magnetic flux annihilation) in the earth magnetosphere [4]. Presence of dust opens a possibility of magnetic field convection and reconnection on astrophysical scale [5].

Recently the Hall physics was applied to consideration of the magnetic field dynamics in the crust of a neutron star, where the magnetic field is frozen into electrons but ions are motionless [6].

The Hall electric field may be important in the circumstellar disks pictured by Hubble ST around young stars created from clumps of molecular cloud [7]. Typical disk has a radius of a hundreds AU, a thickness of tens AU, and is an environment where planets are formed. In the process of accretion under gravity the disk matter begins to rotate and a strong magnetic field should formed in the disks (dynamo effect). Disks and jets propagating along the axis of the disk are very remarkable phenomena. We will present a new possible explanation of these phenomena based on the Hall physics.

For simplicity we will consider the dynamo in usual electron-ion plasma were the Hall physics applies if the time scale of the problem is much less then the ion cyclotron period. Later we will show how to transfer the results of our consideration to the astrophysical dusty plasma.

## 2. Basic equation for the Hall plasma

Let us consider a simple geometry – plasma disk or cylinder with radius $R$. The magnetic field $B$ initially has only poloidal $r$ and $z$ components. The rotating conductor (plasma) begins to stretch magnetic field lines and toroidal component $B_q$ then appears. The kinematics dynamo problem is described by an equation for the magnetic field frozen in the electron liquid. Ions are moving in $q$ direction with a velocity $V(r,z) = r\Omega(r,z)$.

$$n \frac{\partial}{\partial t}\frac{\vec{B}}{n} + n\vec{V}_e \cdot \vec{\nabla}\frac{\vec{B}}{n} = (\vec{B}\vec{\nabla})\vec{V}_e + \frac{\partial}{\partial z}\eta\frac{\partial \vec{B}}{\partial z} \tag{1}$$

$$\vec{\nabla} \times \vec{B} = \frac{4\pi}{c}\vec{j} = -\frac{4\pi}{c}en(\vec{V}_e - \vec{V}) \tag{2}$$

Eqs. (1-2) in the case of a rotating cylinder of Hall plasma may be rewritten as:

$$\frac{\partial B}{\partial t} = -(r\frac{\partial A}{\partial z}\frac{\partial}{\partial r} - \frac{\partial(rA)}{\partial r}\frac{\partial}{\partial z})\Omega - (r\frac{\partial B}{\partial z}\frac{\partial}{\partial r} - \frac{\partial(rB)}{\partial r}\frac{\partial}{\partial z})\frac{B}{rN} -$$
$$-(r\frac{\partial A}{\partial z}\frac{\partial}{\partial r} - \frac{\partial(rA)}{\partial r}\frac{\partial}{\partial z})\frac{1}{rN}(\frac{\partial^2 A}{\partial z^2} + \frac{1}{r}\frac{\partial}{\partial r}r\frac{\partial A}{\partial r} - \frac{A}{r^2}) + \eta(\frac{\partial^2 B}{\partial z^2} + \frac{1}{r}\frac{\partial}{\partial r}r\frac{\partial B}{\partial r} - \frac{B}{r^2}) \tag{3}$$

$$\frac{\partial A}{\partial t} = -\frac{1}{Nr^2}(r\frac{\partial B}{\partial z}\frac{\partial}{\partial r} - \frac{\partial(rB)}{\partial r}\frac{\partial}{\partial z})rA + \eta(\frac{\partial^2 A}{\partial z^2} + \frac{1}{r}\frac{\partial}{\partial r}r\frac{\partial A}{\partial r} - \frac{A}{r^2}) \tag{4}$$

$$B_r = \frac{\partial A}{\partial z}, B_q = B, B_z = -\frac{\partial(rA)}{r\partial r}$$

We are using normalized values. Here $t \rightarrow t\omega_{ci}$, $(r,z) \rightarrow (r,z)c/\omega_{pi}$, $\Omega \rightarrow \Omega/\omega_{ci}$, $\eta = \nu_{ei}/\omega_{ce}$, $B \rightarrow B/B_0$, $A \rightarrow A/A_0$, $B_0$ and $A_0$ are some amplitudes, $\omega_{pi}= (4\pi n_0 e^2/M)^{1/2}$, $\omega_{ci,e} = eB/c(M,m)$, $\nu_{ei}$ is the rate of electron-ion collisions. $N(r,z) = n(r,z)/n_0$, $n_0$ is a plasma density in some point. In sections 3-5 we will consider the incompressible plasma, $N = 1$.

## 3. Instability of rotating Hall plasma in axial magnetic field

Rotating Hall plasma in axial homogeneous magnetic field, $B_z$ ($A_0 = B_z r/2$) and toroidal magnetic field $B_q = B_q(r)$ may be unstable. Let $\Omega = \Omega(r)$ and small perturbation of $B$ and $A$ proportional to $\exp(-i\omega t + ikz)$, $kR \gg 1$. In linear approximation the eqs. (3,4) may be rewritten as:

$$-i\omega dB = -ikr\frac{\partial \Omega}{\partial r}dA + i\frac{2}{r}kB_q dB - ik^3 B_z dA - k^2 \eta dB \qquad (5)$$

$$-i\omega dA = -ikB_z dB + ik\frac{1}{r}\frac{d(rB_q)}{dr}dA - k^2 \eta dA \qquad (6)$$

and we get a dispersion equation:

$$(\omega + \frac{k}{r}\frac{d(rB_q)}{dr} + ik^2\eta)(\omega + \frac{2k}{r}B_q + ik^2\eta) = k^4 B_z^2 + k^2 B_z r\frac{d\Omega}{dr} \qquad (7)$$

The instability may develop if:

$$-B_z r\frac{d\Omega}{dr} > [\frac{r}{2}\frac{d}{dr}(\frac{B_q}{r})]^2 + k^2 B_z^2 + k^2 \eta^2 \qquad (8)$$

This magneto-rotational instability in the limit $B_q = 0$ was found in papers [8,9]. However just a small $B_q$ can stabilized the rotating Hall plasma. In physical values the criterion (8) can be rewritten as:

$$-\frac{r}{\omega_{ci}}\frac{d\Omega}{dr} > \frac{1}{B_z^2}[\frac{r}{2}\frac{d}{dr}(\frac{B_q}{r})]^2\frac{c^2}{\omega_{pi}^2} + \frac{k^2 c^2}{\omega_{pi}^2}(1 + \frac{\nu_{ei}^2}{\omega_{ce}^2}), \omega_{ci,e} = \frac{eB_z}{c(M,m)} \qquad (9)$$

In order to neglect oscillation of the ions, as we did, the phase velocity of wave should be grater than the Alfven velocity $V_A = B/(4\pi\rho)^{1/2}$ where ρ is the density of matter involved in mass motion. In papers [8,9] was considered the instability of a very low ionized dense gas, as in earth ionosphere or protostellar disks. If ion neutral collision rate $\nu_{in}$ is much grater then $\omega$, than ions are moving with a gas and the above theory of magneto-rotational instability can be used for wave numbers $k > (n_i M/\rho)^{1/2}\omega_{pi}/c$. Finally we are getting the criteria of instability:

$$-\frac{r}{\omega_{ci}}\frac{d\Omega}{dr} > \frac{1}{B_z^2}[\frac{r}{2}\frac{d}{dr}(\frac{B_q}{r})]^2\frac{Mc^2}{4\pi e^2 n_i} + \frac{n_i M}{\rho} \qquad (10)$$

### 4. Dynamo effect in a thin rotating disk of Hall plasma

Let the disk have a thickness $d$ much smaller than the radius $R$. We will find a number of particular nonlinear solutions of eqs. (3,4). In the beginning we will consider a case with a small magnetic diffusion, $\eta \to 0$, and will try to find steady state solution ($\partial(B,A)/\partial t = 0$). In the case of a thin disk in the operator

$$(\frac{\partial^2}{\partial z^2} + \frac{1}{r}\frac{\partial}{\partial r}r\frac{\partial}{\partial r} - \frac{1}{r^2})(B, A) \qquad (11)$$

we can ignore the second and the third terms. Then we can eliminate the dependence of $B$ and $A$ on the radius in eqs. (3,4), if

$$B(r,z) = r^m b(z) = kA(r,z) = kr^m a(z), \Omega = \Omega_0 r^{m-1}$$

For a solid rotating disk $m = 1$, for a Keplerian disk $m = -1/2$. Then eqs. (3,4) can be rewritten as:

$$-(m-1)\Omega_0\frac{\partial a}{\partial z} + 2b\frac{\partial b}{\partial z} - (m-1)\frac{\partial a}{\partial z}\frac{\partial^2 a}{\partial z^2} + (m+1)a\frac{\partial^3 a}{\partial z^3} = 0 \qquad (12)$$

$$a\frac{d}{dz}b - b\frac{d}{dz}a = 0, b = ka \qquad (13)$$

Here $k$ is a constant. Substituting $b = ka$ in (12) we get an equation for $a(z)$. This equation has an integral:

$$\frac{d^2a}{dz^2} + k^2a + \Omega_0 = Ca^{\frac{m-1}{m+1}} \qquad (14)$$

$C$ is a new constant resulting from the integration. In particular, we can chose $C = 0$. A set of nonlinear equations (3,4) has a particular solution like that of a linear equation

$$\frac{d^2a}{dz^2} + k^2a + \Omega_0 = 0 \qquad (15)$$

Wave number $k$ can be defined from the boundary condition. Let us consider a rotating disk of Hall plasma placed in metal box (perfect conductors). Then the axial magnetic field can not change near the metal. Also, magnetic flux of $B_q$ and $B_r$ must conserved:

$$\int_{-d/2}^{d/2} B_q dz = 0, \int_{-d/2}^{d/2} B_r dz = A(\frac{d}{2}) - A(-\frac{d}{2}) = 0$$

So, $a$ is an even function of $z$. If we ignore the seeding $B_z$ field which is necessary to start the dynamo, we should take $a = 0$ at the plates surface. Then we get the set of equations for the steady state dynamo in Hall plasma rotating with the angular velocity $r\boldsymbol{W} \sim r^m$, where the parameters $A_0$ and $k$ are defined.

$$A_0 \cos(\frac{kd}{2}) = k^{-2}\Omega_0, \tan(\frac{kd}{2}) = \frac{kd}{2}, B_r = -A_0 k \sin(kz) r^m,$$
$$B_q = k(A_0 \cos(kz) - k^{-2}\Omega_0)r^m, B_z = -(A_0 \cos(kz) - k^{-2}\Omega_0)(m+1)r^{m-1} \qquad (16)$$

In contrary to MHD kinematics dynamo solution, the magnetic field rotation and slipping in mass motion in Hall MHD results in appearing of relatively small magnetic field:

$$B_q \approx B_r \approx \frac{4\boldsymbol{p}end}{c} r\Omega, B_z \approx \frac{d}{r} B_r \qquad (17)$$

Now we will consider a time dependent problem, but for an opposite case – magnetic diffusion is strong enough. Initial magnetic field has $z$ component only ($B_z = B_0$) and a solid Hall plasma disk is rotating with $\boldsymbol{W} = \boldsymbol{W}_0(z)$. Let $\boldsymbol{h} >> d/R$. Then we can ignore the magnetic field rotation, $\boldsymbol{d}A << B$, because of the strong dissipation of the induced poloidal field (flux $A$), and eqs.(3,4) can be rewritten as one equation for $b$:

$$\frac{\partial b}{\partial t} = b\frac{\partial b}{\partial z} + \frac{\partial \Omega_0}{\partial z} + \boldsymbol{h}\frac{\partial^2 b}{\partial z^2}, b = \frac{B_q}{B_0} \qquad (18)$$

For the case $t, \boldsymbol{h}/d >> 1$ we get the basic equation for MHD kinematics dynamo:

$$\frac{\partial \Omega_0}{\partial z} + \boldsymbol{h}\frac{\partial^2 b}{\partial z^2} = 0 \qquad (19)$$

Maximum of the magnetic field $B_q$ can be estimated in physical values as $B_0R\boldsymbol{W}d/\boldsymbol{h}$. However, since the magnetic Reynolds number $R_m = V_A d /\boldsymbol{h}$ ($V_A^2 = B_0^2/4\boldsymbol{p}nM$) is usually very large, even a small Hall term can change the maximum value of $B_q$. In the Hall case $B_q$ can be estimated as $B_0(R\boldsymbol{W}V_A)^{1/2} (R\boldsymbol{w}_{pi}/c)^{1/2}$ and may be much smaller than that estimated above.

Equation (18) can be solved numerically. Below we will find an analytical solution for one particular case. A thin solid disk of Hall plasma is rotating in motionless thick body of Hall plasma. Initially magnetic field is generated near the surface of the disk and penetrates in the Hall plasma as a shock wave described by Burger's-like equation (18) where $d\mathbf{W}/dz = 0$. The amplitude of the shock is $B_0(RU/Hd)^{1/2}$, the velocity is $V_A(UHd/R)^{1/2}$, the front thickness is $(d/R_m)(R/UHd)^{1/2} = \mathbf{h}/V_A$. We introduced the Hall parameter $H = c/\mathbf{w}_{pi}d$ and the normalized velocity $U = R\mathbf{W}/V_A$. The amplitude and velocity of the waves coming from opposite directions to the middle plane have opposite signs. Finally, a stationary anti-symmetrical step-like structure is established (Fig 1).

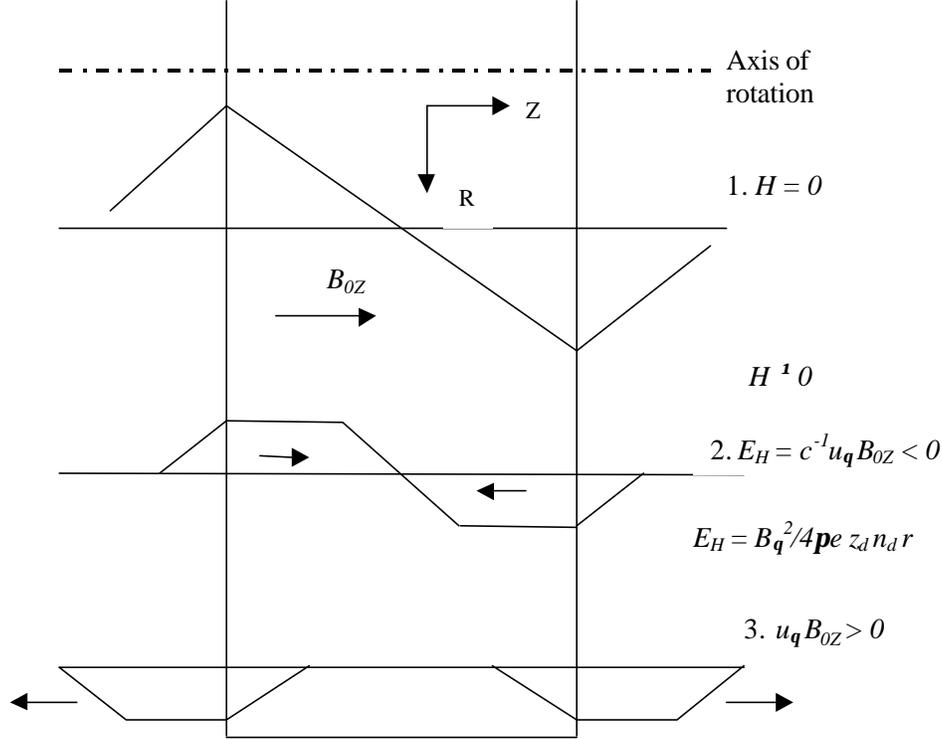

Fig 1. Scheme of magnetic field generation in a thin solid disk of Hall plasma rotating with velocity $u_\mathbf{q}$ in thick Hall plasma body. 1. Hall number $H$ is zero. 2. Hall electric field $E_H$ is saturated near the value $c^{-1}u_\mathbf{q}B_{0Z}$. 3. Disk is rotating in the direction opposite to the case 2. The arrows in cases 2 and 3 are pointing in the direction of the magnetic flux flow.

It is important to mention that in the first considered case we got the dynamo effect in a limit of small magnetic diffusion. Such a dynamo does not dissipate the energy in the steady state limit. The 3-component magnetic field and magnetic energy are accumulated during the transition time. In the second case Ohmic heating is important but much less than in the case of usual MHD kinematics dynamo.

### 5. Electrical jet along a Hall plasma cylinder

We are going to investigate stationary nonlinear waves along a motionless Hall plasma cylinder. Let the plasma density be constant across the cylinder with radius $R$. Let $\mathbf{W} = 0$.

The eqs. (3,4) have the solution $B = rb(z-ut)$, $A = ra(z-ut)$ inside the plasma cylinder. For this case, as for the case of a thin disk, the operator (11) is $\partial^2(B,A)/\partial z^2$ and we get:

$$u\frac{db}{dz} = -2b\frac{db}{dz} - 2a\frac{d^3a}{dz^3} + h\frac{d^2b}{dz^2} \quad (20)$$

$$u\frac{da}{dz} = 2a\frac{d}{dz}b - 2b\frac{d}{dz}a + h\frac{d^2a}{dz^2} \quad (21)$$

This set of equations is similar to eqs. (12,13) and we can find its solution easily in the same way as before. For the slow diffusion case ($\eta \to 0$) the solution is:

$$b = ka - \frac{u}{2}, \frac{d^2a}{dz^2} + k^2 a = C \quad (22)$$

$$a = A_0 \cos(kz) + k^{-2}C, \; B_r = -A_0 k \sin(kz) r,$$
$$B_q = (A_0 k \cos(kz) + k^{-2}C - \frac{u}{2})r, \; B_z = -2A_0 \cos(kz) + 2k^{-2}C \quad (23)$$

$C$ and $A_0$ are constants. This solution is describing nonlinear periodic waves moving with velocity $u$ along the plasma cylinder. In a linear approximation $A_0$ is small and we get a phase velocity of the whistler drift waves (main mode) moving along the plasma cylinder in the magnetic field with only $B_q = rB_0/R$ component $(C = 0)$:

$$u = -cB_0/2\text{p}en_0 R \quad (24)$$

In the limit $kR >> 1$ we get the usual whistler mode, $w = (eB_{z0}/Mc) k^2 c^2 / w_{pi}^2$ if we are chosen $C = B_{z0}/2$.

Now, as in previous section, we will consider a time dependent problem for an opposite case – where magnetic diffusion is strong enough to suppress the poloidal magnetic field appearing due to magnetic field rotation, $A << B$. Then eqs. (20,21) can be rewritten as:

$$\frac{\partial b}{\partial t} = 2b\frac{\partial b}{\partial z} + h\frac{\partial^2 b}{\partial z^2} \quad (25)$$

The toroidal magnetic field $B_q$ is generated near the surface of the disk in the dynamo process and penetrates in the Hall plasma cylinder as a shock wave $B_q = rb(z - ut)$ described by Burger's equation (25). The velocity of the shock wave is given by eq.(24), the front thickness is $-h/u$. An electrical current flows along the surface of the plasma cylinder and returns homogeneously distributed inside the plasma body back to the dynamo. The voltage and radial current induced in dynamo support the electrical jet.

In this section we have shown that energy can be ejected from the disk dynamo in the form of a nonlinear periodic magnetic wave or shock magnetic wave. The moving periodical structures in the jets were observed by Hubble ST. There are a number of numerical simulation in frame of MHD of this phenomenon, see for example Ref. [10,11]. Considered here physics is very different one.

## 6. Equation for the nonlinear structures and waves in the Hall plasma

It is possible to simplify the set of eqs. (3-5) for the steady state solutions, $\partial(B,A)/\P t = 0$. Let $N = N(r,z) = n(r,z)/n_0$ and magnetic diffusion is small, $\eta \to 0$. The general solution of eq. (4) in this case is: $rB = H(rA)$, where $H(rA)$ is an arbitrary function of $rA(r,z)$. The second term in the right part of eq. (3) can be rewritten as:

$$-(r\frac{\partial B}{\partial z}\frac{\partial}{\partial r}-\frac{\partial (rB)}{\partial r}\frac{\partial}{\partial z})\frac{rB}{Nr^2}=-(r\frac{\partial A}{\partial z}\frac{\partial}{\partial r}-\frac{\partial (rA)}{\partial r}\frac{\partial}{\partial z})\frac{dH}{d(rA)}\frac{H}{Nr^2}, rB=H(rA) \qquad (26)$$

Now we see that eq. (3) has an integral:

$$\Omega(r,z)+\frac{H}{Nr^2}\frac{dH}{d(rA)}+\frac{1}{Nr}(\frac{\partial^2 A}{\partial z^2}+\frac{1}{r}\frac{\partial}{\partial r}r\frac{\partial A}{\partial r}-\frac{A}{r^2})=G(rA), N=N(r,z) \qquad (27)$$

where *G* is another arbitrary function of *rA(r,z)*.

We can get similar integrals for the moving structures $B = B(r,z-ut)$, $A = A(r,z-ut)$ if the plasma cylinder is homogeneous along the axis, $N = N(r) = n(r)/n_0$:

$$rB=H(rA)-u\int_0^r Nrdr \qquad (28)$$

$$\Omega(r,z)-u\frac{\int_0^r Nrdr}{Nr^2}\frac{dH}{d(rA)}+\frac{H}{Nr^2}\frac{dH}{d(rA)}+\frac{1}{Nr}(\frac{\partial^2 A}{\partial z^2}+\frac{1}{r}\frac{\partial}{\partial r}r\frac{\partial A}{\partial r}-\frac{A}{r^2})=G(rA) \qquad (29)$$

Here again *H* and *G* are arbitrary functions of *rA(r,z)*.

Eqs. (27) and (28,29) are much simpler than eqs. (3-5), especially for numerical analysis. In particular case of an incompressible plasma, *N = 1*, we are getting easily the solutions of section 4 and 5.

It is necessary to mention that in the case **W**= 0 eq. (27) is looking similar to the Grad-Shafranov equation for the low pressure plasma limit, which was used in Ref. [11] for analysis of the jet formation in the frame of the usual MHD. The difference is in the appearance of a new arbitrary function *G(rA)*. We would like to stress that eq. (27) and the Grad-Shafranov equation are results of very different physics. The Grad-Shafranov equation is describing force-free steady state magnetic field configurations. We consider magnetic field evolution in a Hall plasma where the density distribution is given and magnetic force is balanced by ion inertia as in laboratory experiment with pulsed plasma [3], or the charged dust particle motion is supported by drag force of the hydrogen gas flow as in astrophysical plasma [1]. This could also apply to a solid Hall semiconductor.

### 7. How to apply the Hall dynamo physics to astrophysical dusty plasmas

The Hall effect due to the presence of the dust particles may be important in circumstellar disks and should be taken into account in the problem of astrophysical dynamo. Typically, dust particles are distributed over radius *a* as $N_d(a) \sim 1/a^{3,4}$ and charged. Dust mass density is defined by larger size particles, charge density by smaller ones. Electrons and ions are magnetized, but dust particles are not magnetized because of the drag force from the gas flow and dust moving with the gas.

We can apply the Hall dynamo physics described above to the circumstellar disks. Evaluation of the equations describing the magnetic field dynamics of the astrophysical dusty plasma for the case we are needed can be found in Ref. [5]. In our equations we have to do the following changes:

$$n \to n_d, e \to ez_d, \mathbf{h} \to \mathbf{h}_i, V \to V_d, \mathbf{h}_i = \frac{n_i M c^2}{4\pi e^2 n_i}\frac{1}{(z_d n_d / n_i)^2} \qquad (30)$$

where $V_d$ is the dust mass velocity, $n_d$, $z_d$ are the density and charge of the dust particles, $n_i$ is the ion density, $M$ is the average ion mass, $\eta_i$ is the rate of the momentum transfer between the ions and gas. $n_d$, $z_d$ and $n_i$ were calculated as a function of the molecular density $n_H$, for example, in Ref.[12,13]. For the smallest grain radius $a = 0.03\mu m$ calculated in [12] for molecular clouds at 10 $K$ and $n_H = 10^6 cm^{-3}$ the ion and grain densities are $n_i = 10^{-2} cm^{-3}$, $n_d = 10^{-5} cm^{-3}$, and most of grains are charged by one electron. The "dust inertial scale" $D = (M_d c^2/4\pi e^2 z_d^2 n_d)^{1/2}$ for such a plasma is approximately $10\ AU$. It is comparable to the thickness of the circumstellar disk.

We have considered the dynamo and jet phenomena in a simple two-component plasma in particular because there are very different views of the disks structure in the astrophysical community [14,15]. At least astronomers have seen a lot of dusty plasma near the disk surface, the plasma parameters being close to the above values.

Some groups in the laboratory and ionosphere plasma community have 2-D and now 3-D Hall MHD codes and there is possibility to begin simulation of the considered above objects in dynamics [4,16]. Ours analytical solutions will be useful for benchmarking the codes and understanding the Hall physics.

I thank Dr. J. Drake and Dr. B. Coppi for useful discussion and support.

### References

1. T. W. Hartquist, W. Pilipp and O. Havnes, Astrophysics and Space Science, 246, 243 (1997).
2. Z.-Y., Li, Astrophys. Journal, 526, 806 (1999)
3. A. V. Gordeev, A. S. Kingsep and L. I. Rudakov, Physics Reports, 253, No. 5 (1994).
4. M. A. Shay, J. F. Drake, B. N. Rogers, R. E. Denton, J. Geophys. Res., 106, 3759 (2001).
5. L. I. Rudakov, Physica Scripta, T89, 158 (2001).
6. S. I. Vainshtein, S. M. Chitre, and A. V. Olinto, Phys. Rev. E, 61, 4422 (2000).
7. C.B. O'Dell and V.W. Beckwith, Science, 267, 1355 (1997).
8. M. Wardle, Mon. Not. R. Astr. Soc., 307, 849 (1999).
9. S. A. Balbus and C. Terquem, ApJ, 552, 235 (2001).
10. R. Quyed, R. E. Pudritz and J. M. Stone, Nature, 385, 409 (1997).
11. G. V. Ustyugova, R. V. E. Lovelace, M. M. Romanova, H. Li and S. A. Colgate, ApJ, 541, L21 (2000); R. V. E. Lovelace, H. Li, J. M. Finn and S. A. Colgate, "Pointing outflow and jets from accretion disks", in press.
12. T. Umebayashi and T. Nakano, Mon. Not. R. Astr. Soc., 243, 103 (1990).
13. M. Wardle and C. Ng, Mon. Not. R. Astr. Soc., 303, 239 (1999).
14. M. Wardel, arXiv:astro-ph/9707228 v2 (1997)
15. P. Artymovicz, Space Science Reviews, 92, 69 (2000).
16. J. D. Huba, Phys. Plasma, 2, 2504 (1995).

Rudakov Leonid I., rudakovl@wam.umd.edu